\documentclass[12pt]{article}
\usepackage{epsf}
\usepackage{amsmath}
\usepackage{graphics}
\usepackage{cite}

\renewcommand{\thefootnote}{\#\arabic{footnote}}
\begin{document}

\newcommand{\gtrsim}{ \mathop{}_{\textstyle \sim}^{\textstyle >} }
\newcommand{\lesssim}{ \mathop{}_{\textstyle \sim}^{\textstyle <} }

\newcommand{\rem}[1]{{\bf #1}}

\renewcommand{\thefootnote}{\fnsymbol{footnote}}
\setcounter{footnote}{0}
\begin{titlepage}

\def\thefootnote{\fnsymbol{footnote}}

\begin{center}
\hfill hep-ph/07mmnnn

\vskip .5in
\bigskip
\bigskip
{\Large \bf Flavor Symmetry for Quarks and Leptons}

\vskip .45in

{\bf Paul H. Frampton} 

{\em Department of Physics and Astronomy, University of North Carolina,}

{\em Chapel Hill, NC 27599-3255. }

{\bf Thomas W. Kephart} 

{\em Department of Physics and Astronomy, Vanderbilt University, }

{\em Nashville, TN 37235. }

\end{center}

\vskip .4in

\begin{abstract}
Present data on neutrino masses and mixing favor
the highly symmetric tribimaximal neutrino
mixing matrix which suggests an underlying 
flavor symmetry. A systematic study of non-abelian finite groups
of order $g \leq 31$ reveals that tribimaximal mixing
can be derived not only from the well known tetrahedral flavor symmetry 
$T \equiv A_4$, 
but also by using the binary tetrahedral symmetry 
$T^{'} \equiv SL_2(F_3)$ which does not contain the
tetrahedral group as a subgroup.
$T^{'}$ has the further advantage
 that it can also neatly accommodate  the quark masses
including a heavy  top quark.

\end{abstract}

\end{titlepage}

\renewcommand{\thepage}{\arabic{page}}
\setcounter{page}{1}
\renewcommand{\thefootnote}{\#\arabic{footnote}}

\newpage

\bigskip

In this letter we propose a flavor symmetry for quarks and leptons.

We shall consider only three 
left-handed neutrinos at first. The Majorana mass matrix
${\cal M}$ is a $3 \times 3$ unitary symmetric matrix and without
CP violation has six real parameters.
Let write the diagonal form as ${\bf M} = {\rm diag} (m_1, m_2, m_3)$,
related to the flavor basis ${\cal M}$ by ${\bf M} = U^{T} {\cal M} U$
where $U$ is orthogonal. It is the form of ${\cal M} = U{\bf M}U^{T}$
and $U$ which are the targets of lepton flavor physics. 
One technique for analysis of ${\cal M}$ is to assume texture 
zeros\cite{textures,marfatia, FGY} in ${\cal M}$ and this
gives rise to relationships between the mass eigenvalues $m_i$
and the mixing angles $\theta_{ij}$. For example, it
was shown in \cite{marfatia} that ${\cal M}$ cannot have
as many as three texture zeros out of a possible six but can
have two.
A quite different interesting philosophy is that neutrino masses may arise from
breaking of lorentz invariance\cite{Glashow}. Clearly, a wide range
of approaches is being aimed at the problem.

In the present study we focus on a symmetric texture for ${\cal M}$ 
with only four independent parameters, of the form

\begin{equation}
{\cal M} = \left(
\begin{array}{ccc}
A & B & B \\
B & C & D \\
B & D & C 
\end{array}
\right)
\label{symmetric}
\end{equation}

\noindent The ${\cal M}$ can be reached from a diagonal ${\bf M}$ by the orthogonal
transformation
\begin{equation}
U = \left(
\begin{array}{ccc}
{\rm cos} \theta_{12} & {\rm sin} \theta_{12} & 0 \\
- {\rm sin} \theta_{12}/\sqrt{2} & {\rm cos} \theta_{12} / \sqrt{2} &
- 1/\sqrt{2} \\ 
- {\rm sin} \theta_{12}/\sqrt{2} & {\rm cos} \theta_{12} / \sqrt{2} &
1/\sqrt{2} 
\end{array}
\right)
\label{theta12}
\end{equation}
where one commits to a relationship between $\theta_{12}$
and the four parameters in Eq.(\ref{symmetric}), namely

\begin{equation}
{\rm tan} 2 \theta_{12} = 2 \sqrt{2B (A-C-D)^{-1}}
\end{equation}

\noindent Written in the standard PMNS form\cite{PMNS}
\begin{equation}
U =
\left(
\begin{array}{ccc}
1 & 0 & 0 \\
0 & {\rm cos} \theta_{23} & {\rm sin} \theta_{23} \\
0 & - {\rm sin} \theta_{23} & {\rm cos} \theta_{23} \\
\end{array}
\right)
\left(
\begin{array}{ccc}
{\rm cos} \theta_{13}  & 0 & {\rm sin} \theta_{13} e^{i \delta} \\
0 & 1  & 0 \\
- {\rm sin} \theta_{13} e^{- 1 \delta}  & 0  & {\rm cos} \theta_{13} \\
\end{array}
\right)
\left(
\begin{array}{ccc}
{\rm cos} \theta_{12}  & {\rm sin} \theta_{12}  & 0 \\
- {\rm sin} \theta_{12}  & {\rm cos} \theta_{12}  & 0 \\
0 & 0  & 1 
\end{array}
\right)
\end{equation}
this ansatz requires that $\theta_{23} = \pi/4$ and $\theta_{13} = 0$,
both of which are consistent with present data. These
values of maximal $\theta_{23}$ and vanishing $\theta_{13}$ are
presumably only approximate but departures therefrom, if
they show up in future experiments, could be accommodated by 
higher order corrections.

\noindent To arrive at tribimaximal 
mixing\cite{Cabibbo2,Wolfenstein,Ma,Ma2,HPS,Altarelli}, 
one more parameter $\theta_{12}$
in Eq. (\ref{theta12}) is assigned such that the entries of the
second column are equal, {\it i.e.} ${\rm sin} \theta_{12}
= {\rm cos} \theta_{12}/\sqrt{2}$ which implies that
${\rm tan}^2 \theta_{12} = 1/2$. Experimentally $\theta_{12}$
is non-zero and over $5\sigma$ from a maximal $\pi/4$.
The present value\cite{expt} is ${\rm tan}^2 \theta_{12} = 0.452^{+0.088}_{-0.070}$,
so the tribimaximal value is within
the allowed range. With this identification Eq.(\ref{theta12}) becomes\cite{HPS}

\begin{equation}
U_{HPS} = \left(
\begin{array}{ccc}
\sqrt{2/3} & \sqrt{1/3} & 0 \\
- \sqrt{1/6} & \sqrt{1/3} & - 1/\sqrt{2} \\ 
- \sqrt{1/6} & \sqrt{1/3} & 1/\sqrt{2} 
\end{array}
\right)
\label{HPS}
\end{equation}

\noindent This ensures that the three mixing angles $\theta_{ij}$ are consistent
with present data, although more accurate experiments may require
corrections to $U_{HPS}$. 

The data allow a normal or inverted hierarchy, or a degenerate spectrum.
The tribimaximal mixing, $U_{HPS}$  of Eq.(\ref{HPS}), can accommodate all
three of these neutrino mass patterns and so makes no prediction
in that regard.

The success of $U_{HPS}$ tribimaximal neutrino mixing has precipitated
many studies of its group theoretic basis\cite{Ma,Ma2,Altarelli} and
the tetrahedral group $T \equiv A_4$ has emerged. The present analysis
was prompted by earlier work of the present authors 
in systematically studying {\it all} non-abelian
finite groups of order $g \leq 31$ both for a quark flavor group
\cite{Kephart1} and for orbifold compactification in string theory
\cite{Kephart2}. Our 
question is whether or not $T$ is singled out from these as the neutrino
flavor symmetry?

\begin{center}

{\bf Character Table of $T \equiv A_4$}

$\omega = {\rm exp}(2 \pi i / 3)$

\bigskip
\bigskip

\begin{tabular}{||c||c|c|c|c||}
\hline\hline
 & $1_1$  & $1_2$ & $1_3$ & 3 \\
\hline\hline
$C_1$ & 1 & 1 & 1 & 3 \\
\hline
$C_2$ & 1 & 1 & 1 & -1 \\
\hline
$C_3$ & 1 & $\omega$ & $\omega^2$ & 0 \\
\hline
$C_4$ & 1 & $\omega^2$ & $\omega$ & 0 \\
\hline\hline
\end{tabular}

\bigskip
\bigskip
\bigskip
\bigskip

{\bf Kronecker Products for Irreducible Representations of $T \equiv A_4$}

\bigskip
\bigskip

\begin{tabular}{||c||c|c|c|c||}
\hline\hline
 & $1_1$  & $1_2$ & $1_3$ & 3 \\
\hline\hline
$1_1$ & $1_1$ & $1_2$ & $1_3$ & $3$ \\
\hline
$1_2$ & $1_2$ & $1_3$ & $1_1$ & $3$ \\
\hline
$1_3$ & $1_3$ & $1_1$ & $1_2$ & $3$ \\
\hline
$3$ & $3$ & $3$ & $3$ & $1_1+1_2+1_3+3+3$ \\
\hline\hline
\end{tabular}

\end{center}

\bigskip

The Kronecker products for irreducible representations
for all the fourty-five non-abelian finite groups with order $g\leq31$ are
explicitly tabulated in the Appendix
of \cite{Kephart2}, where the presentation is adapted to a style 
aimed at model builders in theoretical physics rather
than at mathematicians as in \cite{ThomasWood}.

Study of \cite{Kephart2} shows that a promising flavor group is
$\equiv SL_2(F_3)$. The Kronecker products are
identical to those of $T \equiv A_4$ if the doublet representations are omitted
and so the group $SL_2(F_3)$ can reproduce successes of $T$ model building.
The use of $SL_2(F_3)$ as a flavor group first appeared
in \cite{Kephart1} and then analysed in more details in \cite{Lebed}.

$SL_2(F_3)$ has an advantage over $T$ in extension to the quark sector
because the doublets of $SL_2(F_3)$, absent in $T$, allow the implementation
of a $(2+1)$ structure to the three quark families, thus permitting the
third heavy family to be treated differently as espoused
in \cite{Kephart3,Kephart1,Conference}

\begin{center}

{\bf Character Table of $T^{'} \equiv SL_2(F_3)$}

$\omega = {\rm exp}(2 \pi i / 6)$

\begin{tabular}{||c||c|c|c|c|c|c|c||}
\hline\hline
 & $1_1$  & $1_2$ & $1_3$ & $2_1$ & $2_2$ & $2_3$ & $3$ \\
\hline\hline
$C_1$ & $1$ & $1$ & $1$ & $2$ & $2$ & $2$ & $3$ \\  
\hline
$C_2$ & $1$ & $1$ & $1$ & $- 2$ & $- 2$ & $- 2$ & $3$ \\  
\hline
$C_3$ & $1$ & $\omega^2$ & $\omega^4$ & $- 1$ & $\omega^5$ & $\omega$ & $0$ \\
\hline
$C_4$ & $1$ & $\omega^4$ & $\omega^2$ & $- 1$ & $\omega$ & $\omega^5$ & $0$ \\
\hline
$C_5$ & $1$ & $1$ & $1$ & $0$ & $0$ & $0$ & $- 1$  \\
\hline
$C_6$ & $1$ & $\omega^2$ & $\omega^4$ & $ 1$ & $\omega^2$ & $\omega^4$ & $0$ \\
\hline
$C_7$ & $1$ & $\omega^4$ & $\omega^2$ & $1$ & $\omega^4$ & $\omega^2$ & $0$ \\
\hline\hline
\end{tabular}

\bigskip

\newpage

{\bf Kronecker Products for Irreducible Representations of $T^{'} \equiv SL_2(F_3)$}

\bigskip
\bigskip

\begin{tabular}{||c||c|c|c|c|c|c|c||}
\hline\hline
 & $1_1$  & $1_2$ & $1_3$ & $2_1$ & $2_2$ & $2_3$ & $3$ \\
\hline\hline
$1_1$ & $1_1$ & $1_2$ & $1_3$ & $2_1$ & $2_2$ & $2_3$ & $3$  \\
\hline
$1_2$ & $1_2$ & $1_3$ & $1_1$ & $2_2$ & $2_3$ & $2_1$ & $3$  \\
\hline
$1_3$ & $1_3$ & $1_1$ & $1_2$ & $2_3$ & $2_1$ & $2_2$ & $3$  \\
\hline
$2_1$ & $2_1$ & $2_2$ & $2_3$ & $1_1 + 3$ & $1_2 + 3$ & $1_3 + 3$ & $2_1 + 2_2 + 2_3$ \\
\hline
$2_2$ & $2_2$ & $2_3$ & $2_1$ & $1_2 + 3$ & $1_3 + 3$ & $1_1 + 3$ & $2_1 + 2_2 + 2_3$ \\
\hline
$2_3$ & $2_3$ & $2_1$ & $2_2$ & $1_3 + 3$ & $1_1 + 3$ & $1_2 + 3$ & $2_1 + 2_2 + 2_3$ \\
\hline
$3$ & $3$ & $3$ & $3$ & $2_1 + 2_2 + 2_3$ & $2_1 + 2_2 + 2_3$ & $2_1 + 2_2 + 2_3$ &
$1_1 + 1_2 + 1_3 + 3 + 3$ \\
\hline\hline
\end{tabular}

\bigskip
\end{center}

It is important to remark that $T^{'} \equiv SL_2(F_3)$ does not contain 
$T \equiv A_4$
as a subgroup\cite{ThomasWood} so our discussion about quark masses
does not merely extend $T$, but replaces it.

\bigskip

The philosophy used for $SL_2(F_3)$ is reminiscent of much earlier
work in \cite{Conference,Kephart3} where the dicyclic
group $Q_6$ was used with doublets and singlets for the
(1st, 2nd) and (3rd) families to transform as $({\bf 2} + {\bf 1})$
respectively. On the other hand, $Q_6$ is not suited
for tribimaximal neutrino mixing because like all dicyclic groups $Q_{2n}$
it has no triplet representation. Recall that when the
work on $Q_6$ was done, experiments had not established neutrino mixing for
the reason explained in our first paragraph. 

To discuss the model building using $SL_2(F_3)$ we must recall from
the $A_4$ model building \cite{Ma,Ma2,Altarelli} that the leptons
can be assigned\footnote{An alternative assignment is in \cite{Ma3}.}
to singlets and triplets as follows:

\begin{equation}
\begin{array}{cc}
\left. \begin{array}{c} 
\left( \begin{array}{c} \nu_{\tau} \\ \tau^- \end{array} \right)_{L} \\
\left( \begin{array}{c} \nu_{\mu} \\ \mu^- \end{array} \right)_{L} \\
\left( \begin{array}{c} \nu_e \\ e^- \end{array} \right)_{L}  
\end{array} \right\}
3  &
\begin{array}{c} 
\tau^-_{R}~~~ 1_1 \\  \\ \mu^-_{R} ~~~ 1_2 \\
\\ e^-_{R} ~~~ 1_3\\  \end{array}  
\end{array}
\end{equation}

\bigskip
\bigskip
\bigskip
\bigskip

\newpage

\noindent The symmetry breaking pattern of most interest is\cite{ThomasWood}
\begin{equation}
SL_2(F_3) ~~ \longrightarrow ~~ Q ~~ \longrightarrow ~~ Z_4 ~~ \longrightarrow ~~ Z_2 ~~
\longrightarrow ~~ {\rm no ~~ symmetry}
\label{SSB}
\end{equation}
where $Q$ is the quarternionic group
so the first discussion concerns the vacuum alignment will cause the symmetry
to break according to the pattern (\ref{SSB}). Recall that the irreps
of $SL_2(F_3)$ are $1_1, 1_2, 1_3, 2_1, 2_2, 2_3, 3$.

\bigskip

\noindent By study of the character tables of these groups, we can
ascertain the VEVS (Vacuum Expectation Values) which generate the required
spontaneous symmetry breakdown.

\bigskip

\noindent The irreps of Q are $1_1, 1_2, 1_3, 1_4, 2$. Concerning the
crucial first stage of symmetry breaking $SL_2(F_3) 
\longrightarrow Q$, the irreps are related by
\begin{eqnarray}
1_1 & \longrightarrow & 1_1  \nonumber \\
1_2 & \longrightarrow & 1_1 \nonumber \\
1_3 & \longrightarrow & 1_1 \nonumber \\
2_1 & \longrightarrow & 2 \nonumber \\
2_2 & \longrightarrow & 2 \nonumber \\
2_3 & \longrightarrow & 2 \nonumber \\
3 & \longrightarrow & 1_2 + 1_3 + 1_4
\end{eqnarray}
so the breaking requires a VEV in $1_2$ or $1_3$ of $SL_2(F_3)$.
We therefore assign the left-handed quarks, consistent with the 2 + 1
philosophy and the third family treated differently
\cite{Conference, Kephart3} as follows:

\bigskip

\begin{equation}
\begin{array}{c}
\left( \begin{array}{c} t \\ b \end{array} \right)_{L}
~~~ 1_1 \\
\left. \begin{array}{c} \left( \begin{array}{c} c \\ s \end{array} \right)_{L}
\\
\left( \begin{array}{c} u \\ d \end{array} \right)_{L}  \end{array} \right\}
2_1 
\end{array}
\end{equation}

\noindent and similarly the right-handed quarks are assigned as:

\bigskip

\begin{equation}
\begin{array}{c} 
t_{R} ~~~ 1_1  \\ 
\left. \begin{array}{c}c_{R}  \\ u_{R} \end{array} \right\} ~~ 2_2 \\
b_{R} ~~ 1_2 \\ 
\left. \begin{array}{c} s_{R} \\ d_{R} \end{array} \right\} 2_3 
\end{array}
\end{equation}

\newpage

\noindent whereupon the mass matrices are:
\begin{equation}
U = \left( \begin{tabular}{c|c}
$<1_3 + 3>$ & $ <2_1> $ \\  \hline
$<2_3>  $ & $ <1_1>   $
\end{tabular} \right)
\label{UUU}
\end{equation}
\noindent and
\begin{equation}
D = \left( \begin{tabular}{c|c}
$<1_2 + 3>$ & $ <2_3> $ \\  \hline
$<2_2>$ & $ <1_3> $
\end{tabular} \right)
\label{DDD}
\end{equation}

\bigskip
\bigskip

\noindent To implement a hierarchy requires first a VEV to a $SU(2)_L$-doublet
Higgs, $H_{1_1}$ which is in the trivial singlet 
representation of $SL_2(F_3)$, thus giving a
heavy mass to $t$ without breaking $SL_2(F_3)$. This mass
is naturally of order the weak scale $\sim v/\sqrt{2} \sim 175$ GeV.

\bigskip
\bigskip

A VEV to a Higgs $H_{1_2}$ breaks $SL_2(F_3)$ to $Q$ and can give masses
to the $b$ quark and $c$ quarks.
More explicitly, starting from the lagrangian for the $SL_2(F_3)$ model, 
we have the Yukawa terms involving the top, bottom, and charm quarks:

\begin{equation}
Y_t \left( \begin{array}{c} t \\ b \end{array}\right)_{1_1}t_{1_1}H_{1_1}
\end{equation}

\begin{equation}
Y_b \left( \begin{array}{c} t \\ b \end{array}\right)_{1_1}b_{1_2}H_{1_3}
\end{equation}
and

\begin{equation}
Y_c \left[ \left(\begin{array}{c} c \\ s \end{array}\right)_{2_1}c_{_{2_2}}
+
\left(\begin{array}{c} u \\ d \end{array}\right)_{2_1}u_{_{2_2}} \right]
H_{1_3}+Y'_c 
\left[ \left(\begin{array}{c} c \\ s \end{array}\right)_{2_1}c_{{2_2}}
+
\left(\begin{array}{c} u \\ d \end{array}\right)_{2_1}u_{{2_2}}
\right]
H_{3}
\end{equation}
where the subscripts on the quark representations and Higgs doublets 
are the $Sl_2(F_3)$ irreps where they live.

Hence, as stated above, the VEV for the $H_{1_1}$ gives a mass to the top quark, but does not break $SL_2(F_3)$.
The bottom and, in part,  charm quark get their masses from a VEV for the
Higgs $H_{1_3}$ transforming according to the $1_3$ irrep of $SL_2(F_3)$.
Thus giving a VEV to  $H_{1_3}$   gives  masses
to these next heaviest quarks.
This causes the
family group to break from $SL_2(F_3)$ to the quarternionic group $Q$. (As we
shall show below all quarks can acquire a Q invariant mass).
The $b/c$ mass ratio is then simply

\begin{equation}
\frac{m_b}{m_c}=\frac{Y_b}{Y_c}.
\end{equation}

\newpage

\bigskip

\bigskip

The Yukawa couplings $Y_{b,c}$ are free parameters and we can 
therefore get any $m_b/m_c$ ratio we want. The theory is not predictive at this 
stage, but it is at least tunable. We can proceed this way to get the other quark 
mass ratios.

The remaining quark masses are generated from the following Yukawa terms

\begin{equation}
Y_s \left[ \left(\begin{array}{c} c \\ s \end{array}\right)_{2_1}s_{_{2_3}}
+
\left(\begin{array}{c} u \\ d \end{array}\right)_{2_1}d_{_{2_3}} \right]
H_{1_2}
+Y'_s \left[ \left(\begin{array}{c} c \\ s \end{array}\right)_{2_1}s_{{2_3}}
+
\left(\begin{array}{c} u \\ d \end{array}\right)_{2_1}d_{{2_3}} \right]
H_{3}
\end{equation}

\noindent where $s,d $ can get a mass from an $H_{1_2}$ VEV and keep $Q$ unbroken.

Hence, all the quarks can get  masses from $H_{1_2}$ and $H_{1_3}$ VEVs that leave $Q$ unbroken. 
Note that VEVs for $H_3$'s also contribute to masses of the first two 
quark generations,  but not to $t$ and $b$ quark masses. More Higgses
are needed to fill out all the off diagonal terms in the quark mass matrix. 
For instance, an $H_{2_2}$ is needed to avoid a texture zero for $m_{ut}$ if this is desired.

\begin{equation}
Y_{ut} \left( \begin{array}{c} t \\ b \end{array}\right)_{1_1}u_{_{2_2}}H_{_{2_3}}
\end{equation}

However, such a contribution is known phenomenologically to be very small.

However, it is important to observe that the only leptonic mass terms are
from $H_{3}$ VEVs, but under
$SL_2(F_3)\rightarrow Q$ we have 

\begin{equation}
3\rightarrow 1_2+1_3+1_4,
\end{equation}
so a VEV for an $H_3$ breaks
$Q$, and giving multiple $H_3$ VEVs can break $Q$ to $Z_4$, $Z_2$, 
or the trivial group of one element.

If we were to ignore $<H_3>$ and off diagonal
terms the hierarchy of quark masses becomes 
$m_b / (m_c=m_u)  = Y_b / Y_c $ and $ m_s = m_d $ and, because these relations are
unsatisfactory, $<H_3>$ must be significant which is interesting
because, as mentioned above, it controls also the lepton masses. Although five quark
masses remain encoded in parametric Yukawa couplings, the advantage over
the minimal standard model is that the top quark mass is naturally
large. As for all flavor groups, including the present one, the
proliferation of Yukawa couplings $Y_k$ is the principal obstacle
to quantitative calculation of the
quark masses.

\bigskip

To see that a reasonable lowest-order CKM matrix can be achieved 
rewrite $m_i \equiv Y_k v_i$ where $i$ is a $T^{'}$ representation
and $Y_k$ the appropriate Yukawa coupling. Then 
the quark mass matrices in Eq.(\ref{UUU}) amd Eq.(\ref{DDD})
must be diagonalized. We work for simplicity
in the limit
\begin{equation} 
 m^U_{2_1}=m^U_{2_3}=m^D_{2_2}=m^D_{2_3}=0
\end{equation}
which corresponds to taking $V_{ub}=V_{cb}=V_{td}=V_{ts}=0$ 
in the CKM matrix. 
In this case all we need to do is diagonalize 
the $(2\times 2)$ sub-matrices, 
\begin{eqnarray} 
  \tilde{\cal M}_{U} 
  &=& \left[ 
         \begin{array}{c|c} 
           i m^U_{3} & m^U_{1_3} + \frac{m_{3}^Ue^{-i\frac{\pi}{4}}}{\sqrt{2}} \\ 
              \hline
              - m^U_{1_3} + \frac{m_{3}^Ue^{-i\frac{\pi}{4}}}{\sqrt{2}} & m_3^U  
         \end{array}
     \right] 
     \,, \label{sub-MU}\\
   \tilde{\cal M}_{D} 
  &=& \left[ 
         \begin{array}{c|c} 
           i m^D_{3} & m^D_{1_2} +  \frac{m_{3}^D e^{-i\frac{\pi}{4}}}{\sqrt{2}} \\ 
              \hline
              - m^D_{1_2} +  \frac{m_{3}^De^{-i\frac{\pi}{4}}}{\sqrt{2}} & m_3^D 
         \end{array}
     \right] 
     \,. \label{sub-MD}
\end{eqnarray}
We begin by diagonalizing $\tilde{\cal M}_U$, which 
can be achieved by introducing two unitary matrices $U_L,$ and $U_R$ 
satisfying  
\begin{equation} 
  U_L^\dagger \cdot \tilde{\cal M}_U \cdot U_R 
  = 
 \left[ 
\begin{array}{cc} 
   m_u & 0 \\ 
    0& m_c 
 \end{array} 
 \right] 
 \,, 
\end{equation}
or equivalently, 
\begin{equation} 
  U_L^\dagger \cdot \left( \tilde{\cal M}_U \tilde{\cal M}_U^\dagger \right) \cdot U_L 
  = 
 \left[ 
\begin{array}{cc} 
   m_u^2 & 0 \\ 
    0& m_c^2 
 \end{array} 
 \right] 
 \,, \label{diagonalization}
\end{equation}
where 
 \begin{eqnarray} 
  \tilde{\cal M}_{U} \tilde{\cal M}_{U}^\dagger
  &=& \left[ 
         \begin{array}{c|c} 
           \frac{3}{2} [m_3^U]^2 + [m_{1_3}^U]^2 + m_{1_3}^U m_3^U  & 
\sqrt{2} m_3^U m_{1_3}^U e^{-i\frac{\pi}{4}} \\ 
              \hline
             \sqrt{2} m_3^U m_{1_3}^U e^{i\frac{\pi}{4}} &  \frac{3}{2} [m_3^U]^2 + [m_{1_3}^U]
^2 - m_{1_3}^U m_3^U
         \end{array}
     \right] 
     \,. \label{sub-squareMU}
\end{eqnarray}
 From Eqs.(\ref{sub-squareMU}) and (\ref{diagonalization}) 
the eigenvalues are calculated: 
\begin{equation} 
  m_{u, c}^2 = \frac{(3[m_3^U]^2 + 2 [m_{1_3}^U]^2) \mp 2\sqrt{3} m_3^U m_{1_3}^U }{2} 
  \,. 
\end{equation}
which indicates how the quark mass spectrum can be successfully accommodated. The unitary matrix $U_L$ takes the form: 
\begin{equation} 
 U_L = \frac{1}{\sqrt{1+A^2}} 
   \left[ 
    \begin{array}{cc} 
      1 & e^{-i\frac{\pi}{4}} A \\ 
      - e^{i \frac{\pi}{4}} A & 1 
    \end{array}
   \right]
   \,, \label{UL}
\end{equation}
with $A=(\sqrt{3}+1)/\sqrt{2}$. 
We note that $U_L$ is independent of quark masses.

In the down-sector, 
since the mass matrix $\tilde{\cal M}_D$ takes the same form 
as $\tilde{\cal M}_U$, a matrix $(\tilde{\cal M}_D \tilde{\cal M}_D^\dagger)$ 
is diagonalized by using the same unitary matrix $U_L$ of Eq.(\ref{UL}). 
Hence we reach the result for this special case  
\begin{equation} 
  V_{\rm CKM} = \left[ 
      \begin{array}{c|c}
     U_L^\dagger D_L & 0 \\ 
      \hline 
     0 & 1 
       \end{array}      
 \right]      
 = \left[ 
     \begin{array} {c|c}
       U_L^\dagger U_L & 0 \\ 
       \hline 
       0 & 1 
       \end{array} 
       \right] 
= {\bf 1} 
\,.
\end{equation}
which is an acceptable first approximation to the CKM matrix.

\bigskip
 
For the neutrinos, as in earlier work on $T$ \cite{Ma},
the masses are not uniquely predicted but the tribimaximal
mixing angles \cite{HPS} are.  All these three neutrino mixing angles are 
consistent with existing measurements.

The Higgs VEVs with a commonality between quarks and leptons are
in the $H_3$ of $T^{'} \equiv SL_2(F_3)$ which has a simple decomposition
under the quarternionic subgroup $Q$ which is likely to play a key role
in the goal of linking lepton masses with quark masses.

\bigskip

In summary, while $T \equiv A_4$ is one candidate for a lepton flavor group which naturally
gives rise to tribimaximal mixing, it is not 
unique among the non abelian finite groups in this regard. 
The choice $T^{'} \equiv SL_2(F_3)$, also known as the binary tetrahedral group
\cite{coxeter}, satisfies the requirement
equally well, and because it has doublet representations
can thereby begin to accommodate the quark mass spectrum, particularly the
anomalously heavy third family \cite{X}.
If our choice is the correct flavor symmetry, 
it remains to understand why Nature chooses the triplet representations
for leptons and the doublet representations for
quarks. Quantitative results for masses will require a relationship
between the Yukawa parameters from our proposed symmetry.

\bigskip

\bigskip
\bigskip

\begin{center}

{\bf Acknowledgements}

\end{center}

\bigskip
TWK thanks the Aspen Center for Physics for hospitality while this research was in progress.
This work was supported in part by the
U.S. Department of Energy under Grants No. DE-FG02-06ER41418 (PHF) and DE-FG05-85ER40226 (TWK).

\end{document}